\documentstyle[aas2pp4]{article} 

\lefthead{Y. Sofue, et al.}
\righthead{Rotation Curves in the \ha\ and
[NII] Line Emissions}

\begin{document}

\def\cen{\centerline}
\def\Msun{M_{\odot \hskip-5.2pt \bullet}}
\def\kms{km s$^{-1}$}
\def\kms{km s$^{-1}$}
\def\co{ CO ($J=1-0$) }
\def\Sobs{\sigma_{\rm obs}}
\def\Scor{\sigma_{\rm cor}}
\def\SISM{\sigma_{\rm ISM}}
\def\Vrot{V_{\rm rot}}
\def\Vt{V_{\rm t}}
\def\deg{$^\circ$}
\def\vlsr{$V_{\rm LSR}$}
\def\vrot{$V_{\rm rot}$}
\def\Vrot{V_{\rm rot}}
\def\ha{H$\alpha$}
\def\Ha{H$\alpha$}
\def\ta{$T^*_{\rm A}$}
\def\Ta{T^*_{\rm A}}
\def\tb{$T_{\rm B}$}

\title{Nuclear-to-Disk Rotation Curves of Galaxies in the H$\alpha$
and [NII] Emission Lines}

\author{Yoshiaki SOFUE\altaffilmark{1}, Akihiko TOMITA\altaffilmark{2},
Yoshinori TUTUI\altaffilmark{1}, Mareki HONMA\altaffilmark{1},
and Yoichi TAKEDA\altaffilmark{1}}
\affil{1. Institute of Astronomy, University of Tokyo, Mitaka,
Tokyo 181-8588, Japan}
\affil{2. Faculty of Education, Wakayama University, Wakayama 640-8510,
Japan}

\begin{abstract}
We have obtained optical CCD spectroscopy along the major axes of 22
nearby spiral galaxies of Sb and Sc types in order to analyze their
rotation curves.
By subtracting the stellar continuum emission, we have obtained position
velocity (PV) diagrams of the \ha\ and [NII] lines.
We point out that the \ha\ line is often superposed by a broad stellar
absorption feature (Balmer wind) in the nuclear regions,
and, therefore, the [NII] line is a better tracer of kinematics in the
central a few hundred pc regions.
By applying the envelope-tracing technique to the \ha\ and [NII] PV diagrams,
we have derived  nucleus-to-disk rotation curves of the observed galaxies.
The rotation curves rise steeply within the central a few hundred
parsecs, indicating rapidly rotating nuclear disk and mass concentration
near the nucleus.
\end{abstract}


\section{Introduction}

Rotation curves have been derived by optical and HI-line observations
in order to study the mass distribution in the disk and
halo of spiral galaxies (Rubin et al 1980, 1982; \cite{bos81};
Kent 1986, 1987; \cite{mat92,mat96,per95,per96}) and of the
Galaxy (\cite{cle85,hon97}).
Current optical observations for spiral galaxies
have been  devoted mainly to
the study of massive halo and disk region, and, therefore,
the spectra were often over-exposed
in the central regions due to the bright stellar bulge light,
not necessarily providing accurate data for the nuclear rotation curve.
HI observations are useful for investigating the
halo and total mass for its extended distribution, while
they are also not necessarily useful to study the  central
few  kpc region because of its deficiency.

Recently, we have shown that the CO line emission will be useful to
derive nuclear rotation curves because of the high central concentration
of CO gas and negligible extinction, as well as for the available high
angular resolution
(Sofue 1996, 1997; Sofue et al 1997, 1998).
We have found that the CO rotation curves show steep nuclear rise
within the central a few hundred pc, indicating a mass concentration
around the nucleus.
Since the CO gas usually coexists  with HII regions,
\ha-line spectroscopy will be also useful to
derive nuclear rotation curves, if the bulge light can be
adequately subtracted and extinction is small enough.
In fact, Rubin et al (1997) have obtained extensive
\ha\ spectroscopy for Virgo spirals,
and have shown a steep rise of the central rotation curves
for many galaxies, indicating a mass concentration.

In order to derive detailed nuclear rotation curves for more number
of galaxies and to examine if the steep nuclear rise of rotation
is universal in spiral galaxies, we have performed optical
spectroscopy of 22 spiral galaxies of Sb and Sc types
using a 188-cm reflector.

\section{Observations}

Observed galaxies are listed in Table 1 with their parameters.
The CCD spectroscopic observations were made using the
Cassegrain spectrometer equipped on
the 188-cm reflector at the Okayama Astrophysical Observatory.
The observations were carried out on 1997 February 28 -- March 6,
and October 2 -- 8.
The original slit length was 5$'$ and was put across the nuclei
and slit position angles were set along the major axes of galaxies.
The Photometrics CCD chip had $512 \times 512$ pixels.
The spatial resolution was $0''.75$~pixel$^{-1}$, and we have bin-ed every
2 pixels in order to increase the signal-to-noise ratio
($1''.5$~bin$^{-1}$).
The spectral dispersion was 0.767\AA
(velocity spacing of 34.9 \kms) per pixel.
The slit width in the wavelength direction was 0.30 mm
in order to get the best combination of effective spectral resolution,
and signal-to-noise ratio.
Considering the slit width, the effective spectral resolution was 1.92A,
which corresponds to a velocity resolution of 87.1 \kms.
This velocity resolution enabled us to determine the radial velocity
at an error of about 9 \kms\ ($\sim 0.1$ times the resolution),
if the signal-to-noise ratio is sufficiently high (e.g., $>$5 rms noise).

The exposure time was 1000 seconds per one spectrum, and we took
three spectra for each galaxy, and finally combined them into one spectrum.
Therefore, the total exposure time for the final one frame was 3000 seconds.
We have subtracted the stellar bulge continuum emission
by applying two methods:
(1)the background-filtering technique (\cite{sof79}),
(2) and a median-average subtraction in the wavelength  direction,
In the second method, we first obtained median value in  the $\lambda$
(wavelength) direction at each position (pixel) on the slit (distance
along the major axis). The median value in the $\lambda$ direction
gives an averaged value of the CCD counts of the
stellar continuum emission, because the $\lambda$ pixel number including line
emissions is much less than that from the continuum part.
Both the methods, (1) and (2), gave almost the same results, while the latter
was found to be better, when the emission lines are wide and strong,
particularly in the central regions.
Threfore, we adopt the results from the median-average subtraction method.
We have, then, subtracted the atmospheric lines, which showed no
variation in intensity and wavelength along the slit.

\cen{--- TABLE 1 ---}
\placetable{table1}

\section{Results}

\subsection{Position-Velocity Diagrams}

The upper two panels of Figure 1 show the obtained position-velocity
diagrams in the \ha\ and [NII] lines in a contour form,
where the abscissa is the relative position in seconds of arc
along the major axis.
The velocity is heliocentric, referred to the
\ha\ 6562.8 \AA\ and [NII] 6583.4\AA\ lines
The contours are drawn in logarithm of base 2, with the lowest contour
level set at 3 counts. That is, each of the contours correspond to
intensity levels of 3, 6, 12, 24, 48, ... detector counts, respectively.

\cen{--- Fig. 1 ---}
\placefigure{fig1}

The vertically-drawn dotted line indicates the position of
the galaxy center, which is defined as being the intensity-peak
in the light of bulge-continuum:
We obtained a continuum-intensity distribution,
i.e., intensity (counts) as a function of position (arcsec),
at each wavelength.
We thus determined the spatial position
corresponding to the continuum-peak at this wavelength.
Repeating this procedure at all wavelength eventually yielded
an ensemble of data, all of which may be the candidates of
the central position of the galaxy. Constructing a histogram
based on these data, we could reasonably establish the position of
the galaxy-center as the most frequent value.
 The length of the horizontal short bar
crossing this center-line shows the actual (physical) scale on the galaxy
corresponding to 2kpc (i.e., extension of $\pm 1$
kpc from the galaxy center),
which was evaluated by using the redshift-based distance ($D$) given
in table 1.

Except for a few cases, the PV diagrams are characterized by
two components:
a bright nuclear component with a large velocity gradient and/or
dispersion, indicating a rapidly rotating nuclear disk,
and a flat component showing a nearly-constant rotation in the disk
and outer region.

\subsection{Line Profiles}

The lowest panels of Figure 1
show the spectral profiles of the \ha\ and [NII] lines
at individual positions on the major axis, drawn in the
form of three-dimensional plot.
The abscissa is the wavelength (radial velocity) and the ordinate
is the distance along the major axis.
For the purpose of background-smoothing, that helps to clarify
the emission feature of present interest, any value
(irrespective of positive or negative) below the threshold level
of 3 counts was intentionally set to zero.
These plots can be used to compare the line profiles of the \ha\
and [NII] lines, whose shapes show complicated variations near to
the nucleus.

In the disk regions, the \ha-to-[NII] line intensity ratio (HNR)
is usually equal to about three, which is typical for HII regions
(e.g., \cite{oster89}), and the two lines show almost the same profiles.
The \ha\ line profile of the nuclear component often shows a
double-horn feature, representing an unresolved rotating disk.
However, the nuclear \ha\ emission is sometimes superposed by a
broad stellar absorption lines, likely due to the Balmer wing of A-type
stars, such as in the case of NGC 3521, in which no
emission feature is recognized any more, being merged by a broad, strong
absorption.
 
Some galaxies like NGC 5033
shows superposition of a broad Seyfert emission feature on the \ha\ line.
Thus, the \ha\ PV diagrams and line profiles very close to the nucleus are
often deformed from the pure emission features, which may cause pseudo
kinematical properties.
On the other hand,
the [NII] line is not affected by such a stellar absorption
feature or by the Seyfert broad emission,
though the signal-to-noise ratio is poorer than the \ha\ line
by a factor of a few.
Considering that the reconstruction processes of the \ha\ P-V diagram
contains uncertainty,
it may be more safe to use [NII] line in the very central regions
where the lines have enough signal-to-noise ratios.

\section{Rotation Curves}

\subsection{Velocity curves from the centroid wavelengths}

First, we derive velocity curves using the centroid of wavelengths
of H-$\alpha$ 6563 and [N II] 6583 emission lines at each spatial position.
Fig. 2 shows the results, where the
closed circles indicate data for H-alpha 6563,
and open circles for [N II] 6583 line.
The vertical dotted-line and the horizontal bar
have the same meaning as in the case of contour diagrams in figure 1.

We define radial velocitis as follows:
In a one-dimensional spectrum $s_m(n)$ at each spatial position $m$,
with $s$ being the intensity, $n$ wavelength or pixel number in the direction
of dispersion, the position of emission peak $n_0$,
which is defined as that satisfying the following conditions
simultaneously, was detected automatically by using a small program.
(i)   $ s(n_0-2) < s(n_0-1)$   and $ s(n_0-1) < s(n_0)$
     and $s(n_0) > s(n_0+1)$   and  $s(n_0+1) > s(n_0+2)$.
(ii) $s(n_0)$ is above the (assumed) threshold level of 5 counts.
Such determined peak position can be directly converted to
the (peak) radial velocity.
However, the resulting plots were unsatisfactory
because of the finite size of pixels; i.e., something like
discontinuously assembled line-segments, rather than a continuous curve.

We decided, therefore, to use the centroid-wavelength instead of
the peak-wavelength:
Once the peak wavelength position, $n_0$, is established,
it is easy to find the two wavelength positions corresponding to
the half-maximum intensity $s(n_0)/2$ by an appropriate interpolation.
Let these two values be $n_-$ and $n_+$
(the inequality of $n_- < n_0 < n_+$ holds),
respectively.
Then, the centroid wavelength of the emission line
is defined as the mean of $n_-$ and $n_+$, i.e., $((n_-)+(n_+))/2$ ].
The rotation curve constructed by using the radial velocity
corresponding to such mean-wavelength at half-maximum turned out
to be more continuous and natural to eyes, which may thus be
more preferable to the peak-velocity curve.

\cen{--- Fig. 2 ---}
\placefigure{fig2}

\subsection{Rotation curves from envelope-tracing method}

We next derive rotation curves from the PV diagrams
by applying the envelope-tracing method (Sofue 1996).
The velocity dispersion of the interstellar gas
($\SISM$) and the velocity resolution of observations
($\Sobs$) are corrected for as the following:
The correction for the finite velocity resolution
$\Sobs=0.5 {\rm FWHM}$ (full width of half maximum) is given by
$$\Scor=\Sobs {\rm exp}(-[d/\Sobs]^2). \eqno(1) $$
Here, $d$  is the velocity difference  between the
maximum-intensity velocity near the profile edge,
and the half-maximum velocity (half width of
the intensity slope at the velocity edge). Equation (1) implies that,
if the original velocity profile of the source is sharp enough,
the observed profile becomes the telescope velocity profile, so that
the correction is equal to the velocity resolution ($\Scor=\Sobs$).
On the other hand, if the velocity profile of the source is extended
largely ($d \gg \Sobs$), the correction becomes negligible
($\Scor \sim 0$) and the half-maximum velocity gives the rotation
velocity.

The terminal velocity $\Vt$ is defined by a velocity at which
the intensity becomes equal to
$$I_{\rm t}=0.5 I_{\rm max} \eqno(2)$$
on the PV diagrams.
Then, the rotation velocity is estimated by
$$\Vrot=[\Vt-(\Scor^2 + \SISM^2)^{1/2}]/{\rm sin}~i.\eqno(3)$$
For the \ha\ and [NII] lines in the present observations,
we take  $\SISM \sim 7$ \kms and $\Sobs=87/2=43.5$ \kms.
The error in estimating the rotation velocity is
$\pm 10$ \kms\ for regions with sufficient signal-to-noise ratio, e.g.,
in the central regions and the disks.
The obtained rotation curves for the \ha\ and [NII] data are shown in
Figures 3 with the abscissa in kpc.
The inserted horizontal bar indicates the angular scale of 30$''$ on the sky.
The distances to the galaxies have been adopted from Table 1, column 9 
(calculated by $H_0=75$ \kms), except for galaxies with redshifts
smaller than 200 \kms, for which Tully-Fisher distances have been
adopted from column 10 of Table 1.
Figure 4 plots all the obtained rotation curves in the same panel.

\cen{--- Fig. 3 ---}
\placefigure{fig3} 

\cen{--- Fig. 4 ---}
\placefigure{fig4} 

\section{Description of Individual Galaxies}

{\bf NGC 1003}:  This is a highly-tilted Sc galaxy, having a slow rotation.
The nuclear component is not visible. The PV diagram shows a strong
asymmetry, with a bright HII region being present at 20$''$ east of the
nucleus.

{\bf NGC 1417}: The data are noisy, so that the PV diagram shows only a
flat disk component in both sides of the nucleus. No nuclear component
is visible in the present data, probaby for the poor data quality.

{\bf UGC 03691}: No clear nuclear component is visible in the PV diagram.
The rotation curve rises rapidly in a rigid-body fashin, reaching a
flat rotation at $R>\sim 20''$.

{\bf NGC 2403}: This Sc galaxy has a  morphology similar to the nearby Sc
galaxy M33, showing amorphous spiral features.
The PV diagram shows an almost rigid-body rotation at $R<50''$,
with a larger gradient in the central $\pm 10''$.
The rotation curve is similar to that of M33.
The disk rotation is almost flat, but it still appears to increase grdually
toward the observed edge.
The \ha-to-[NII] line intensity ratio (HNR) is as large as $\sim 5$ at the
nucleus, slightly larger than the normal value of about three for HII regions.

{\bf NGC 2590}: This Sbc galaxy shows a high-velocity nuclear disk, with a
rapidly increasing rotation at $R<2''$ (600 pc), followed by a flat rotation
until $R \sim 50''$ (16 kpc).
The nuclear \ha\ disk is brighter than the disk part (Fig. 2), and the
rotation property is similar to that of NGC 4527 (Sofue et al 1998).
The HNR at the nucleus is close to the value of normal HII regions.
Since the distance is as large as 64.5 Mpc, no detail can be seen from the
present data.

{\bf NGC 2708}: The  HNR both in the disk and nuclear region
is equal to about three, typical for HII regions.
The PV diagram shows a typical nuclear rise, followed by a flat rotation in
the disk.

{\bf NGC 2841}: The nuclear component has an inverse HNR, as small
as $\sim 0.3$, and the \ha-line width is smaller than [NII].
This may be due either to a superposed broad \ha\ absorption, or to stronger
emission in [NII] such as by higher-temperature circum nuclear gas.
The [NII] nuclear component shows a high concentration of the gas.
The nuclear PV diagram is slightly tilted in the sense of the galactic
rotation, indicating a compact, rapidly rotating gas.
The PV behavior of the disk part is flat both in the \ha\ and [NII] lines,
with a normal HNR value for HII regions.
The rotation velocity of this galaxy is as high as $\sim 300$ \kms\ in the
disk, and reaches almost 350 \kms\ in the nuclear disk.

{\bf NGC 2903}:  The \ha\ and [NII] lines show an almost identical behavior
both in the PV and  intensity plots. The HNR is also normal as for HII regions.
The nuclear component shows a tilted double peaks in the PV diagram,
indicating a rotating ring of radius 4$''$ comprising HII regions.
After decreasing to a minimum at $R\sim 25''$, the disk rotation gradually
increases until $R\sim 50''$. Then, it increases suddenly, with a step,
to a maximum at $R\sim 70''$, followed by a flat rotation.
This step-like rotation may be related to its barred structure.

{\bf NGC 3198}:  This is a highly titlted SBc galaxy. The nuclear component
shows a rapid increase of rotation within 2$''$, while its rotation velocity
is as small as 50 \kms. The rotation velocity, then, increases smoothly to a
maximum at $R\sim 60''$, followed by a flat rotation.
HNR value is for HII regions, and remains nearly constant through the disk and
nuclear component.

{\bf NGC 3495}: The nuclear component is weak both in \ha\ and [NII].
The PV diagrams show a rigid-body behavior at $R<\sim 50''$, followed by a
flat part in the outer disk.
This apparently rigid-body rotation may be due to interstellar extinction
by the disk at a high inclination.

{\bf NGC 3521}: This highly tilted Sbc galaxy shows a strong and broad
\ha\ absorption around the nucleus.
The absorption feature is as wide as $\sim 20$A ($\sim 1000$ \kms), and
extended for about $10''$, but shows little rotation.
On the other hand, the [NII] line suffurs from no absorption, and can
better be used for tracing the kinematics in the PV diagram.
The nuclear component in the [NII] line indicates a rapidly rising rotation
curve of the nuclear disk.
The disk part shows a nearly flat rotation both in the \ha\ and [NII] lines.

For its very slow, or little, rotation and extended feature, the \ha\
absorption component is most likely caused by a stellar absorption line
of the central bulge stars.
Such a broad absorption line is well known for A type stars as the damped
wing in the Balmar lines due to Stark broadening.
In order for such a bright A type star cluster be present, a large-scale
starburst may have happend within the last $10^9$ yrs
(e.g., \cite{dressler83}).

{\bf NGC 3672}: The nuclear component shows a compact, slowly rotating core,
followed by a rigid-rotation feature of the PV diagram.
The rotation velocity attains a maximum at $R\sim 35''$, and then, it is
almost flat.
The HNR is nearly constant with a value for HII regions.

{\bf NGC 4062}: A weak nuclear component is visible both in the \ha\ and [NII]
lines, indicating a slowly rotating compact disk.
The rotation velocity reaches its maximum at $R\sim 15''$, beyond which the
rotation is nearly flat.

{\bf NGC 4321}: This Sc galaxy has a high concentration of \ha\ emission near
to the nucleus, and the PV diagram has a tilted double-horn feature,
representing a rotating ring of HII regions.
However, the central region within a few arcsec is
associated with a broader
[NII] feature than \ha, representing a slightly titlted PV ridge,
suggesting a more compact, rapidly rotating component with a steeply
rising rotation curve.
Except for this [NII] steep component, the HNR in the nuclear and
disk components is normal for usual HII regions.

{\bf NGC 4448}: A compact nuclear component is found in the \ha\ and [NII]
PV diagrams, indicating a steeply rising rotation velocity.
The \ha\ line appears to be superposed by a broad absorption feature,
which may result in the narrower velocity dispersion of \ha\ than [NII].
The [NII] nucear component shows a faster rotating feature than \ha.
The disk part is rather weak, with both \ha\ and
[NII] lines showing a flat rotation.

{\bf NGC 4527}: Double peaked nuclear PV feature is clearly visible both
in the \ha\ and [NII] line emissions, indicating a rapidly rotating disk
of $R\sim 4''$.
The [NII] PV diagrams shows a broader, and more steeply rising component
of rotation near to the nucleus within the central $\sim 2''$.
The rotation veocity, then, decreases to a minimum at $R\sim 15''$, and
increases to a flat part from $R>25''$.
A detailed analysis of the PV diagrams and CO line data is presented in
Sofue et al (1998).

{\bf NGC 4565}: This is an edge-on galaxy of Sb type.
The nuclear component is weak in the \ha\ emission, while
rather strong in [NII].
This may be due to a superposed broad stellar absorption feature of the
\ha\ damped Balmer wing due to the nuclear stellar disk and bulge stars.
The nuclear [NII] feature in the PV diagram indicates a steep rise within
$\sim 2''$, representing a rapidly rotating nuclear disk.
A rigidly rotating PV feature runs across this nuclear PV feature both in
the \ha\ and [NII] lines, which is due to the outer disk component.
The rotation curve is flat beyond $R \sim 50''$.
Since the interstellar extinction by the disk is significant,
the nuclear rotation curve may not be correctly shown up in the
present wave length: the nuclear rise may be much sharper.
In fact,  the CO-line data, which does not suffer from the
extinction, shows a high velocity in the central 5 kpc, where the
rotation velocity remains almost at around 250 \kms\ (Sofue 1997).

{\bf NGC 4569}: The [NII] line is stronger in the nuclear PV component,
which shows a steep rise of rotation within $\sim 1''$, though the
detail is not resolved in the present data.
The rotation velocity then attains a minimum at $\sim 15''$, and
increases to a flat part beyond $R \sim 30''$.

{\bf NGC 4605}: No nuclear component is visible.
The rotation appears to increase in a rigid-body fashion.

{\bf NGC 5033}: The nuclear component is superposed by a broad \ha\ emission
feature from the unresolved nucleus, as wide as $\sim 1000$ \kms.
The [NII] line does not suffer from this nucleus emission, and better
trace the kinematics.
The [NII] PV diagram shows a steep rise of rotation within $\sim 2''$,
and then a minimu at $10''$, followed by a flat rotation at $R>15''$.

{\bf NGC 5907}: This is an almost perfect edge-on Sc galaxy.
The nuclear component is not visible because of the extinction by the
foreground disk.
The disk component shows a rigidly rotating PV feature at $R<60''$,
beyond which the rotation is flat.
This rigid-body-like behavior occurs, because  we are looking at a
ring  in the disk, where the optical depth along the lines of
sight becomes nearly equal to unity.
The CO line data shows a sharply rising rotation curve (Sofue 1997).

{\bf NGC 7331}: The nuclear component is hardly visible in the \ha\ line,
while it is evident in [NII], showing a rapidly rising rotation at $R<2''$.
The rotation of the nuclear disk is rather slow, $\sim 100$ \kms.
The disk part is clearly traced in the \ha\ line, reaching a maximum
at $R\sim 35''$, beyond which the rotation is flat.

\section{Discussion}

\subsection{H$\alpha$ vs [NII] Lines}

The \ha-to-[NII] line intensity ratio (HNR) is equal to
about three for normal HII regions (e.g., \cite{oster89}).
This applies to the disk regions of the galaxies observed in this paper,
where both the lines have shown almost identcal profiles with a constant
line ratio.
However, the \ha\ line profiles in the nuclear regions of some galaxies
(e.g., NGC 3521) were found to be superposed by a broad stellar
absorption line due to the Balmer wing most likely by A type stars.
In Fig. 5, we compare the \ha\ and [NII] lines of NGC 3521 in a grey scale,
in order to demonstrate how the \ha\ Balmer wing affects the PV diagram
of the nuclear region.
In principle, this will happen to any galaxies, if their nuclear region
contains a considerable number of A type stars such as due to starbursts.
Therefore, in this paper, we have useed the [NII] line for the central
regions in place of \ha\ line.

\cen{--- Fig. 5 ----}
\placefigure{fig5}

The forbidden [NII] line has some advantages for discussing the kinematics
of the nuclear regions:
it is not affected by the stellar absorption feature,
and no particular correction for the difference in
wavelength is necessary when used in place of \ha, since
both the lines suffer from an almost equal amount of interstellar extinction.
The origin of the [NII] line emission is still unclear,
either if it comes from diffuse regions around normal HII regions,
or from a circum-nuclear ionized gas.
In either case, however, its PV behavior will indicates the rotation
of the nuclear gas disk, in so far as the [NII]-emitting gas is
gravitationally bound to the galaxy.
In fact, this will be the case, because the sound velocity of [NII]-emitting
gas is of the order of $\sim 30$ \kms, or the temperature $\sim 10^5$ K,
the gas cannot escape from the galactic center.

\subsection{Comparison with CO-line and Previous \ha\ Data}

Some of the galaxies observed in this paper have been  discussed
in our recent paper on the CO-line rotation curves (Sofue 1996, 1997).
In most cases, the \ha/[NII] rotation curves are consistent with
the CO-line rotation curves.
This is a natural consequence of the fact that the molecular clouds (CO)
and HII regions (\ha-emitting regions) are tightly associated with
each other, if optical extinction and absorption are not significant.
However, in such galaxies like NGC 3521, whose \ha\ line is strongly
affected by the stellar Balmar absorption wing, the \ha\ results
are not necessarily coincident with the CO results.
In such galaxies, [NII] rotation curves are more consistent with the
CO data, which is indeed found to be the case for NGC 3521 (Sofue 1997).
Therefore, [NII] line would be a better tracer of kinematics than \ha-line
for the central regions of galaxies.

The present results are consistent with the recent \ha\
study of rotation curves of Virgo galaxies by Rubin et al (1996),
who have shown that many galaxies have steeply rising rotation
in the center, indicating rapidly rotating nuclear gas disks.
However, the nuclear rotation curves derived from
our CCD data are not necessarily consistent with the
earlier \ha\ observations using photographic plates
(Rubin et al 1982; Mathewson et al 1992; Persic et al 1996),
which usually showed a rigid-body increase
of rotation velocity within a few kpc region.
This is because of the difference in the dynamic range of observations,
particularly in the central regions, where the bright bulge continuum
affect the PV diagrams significantly:
The earlier photographic data were often saturated in the central
bulge regions, and also their interest was more in the outer
flat part of rotation curves, but not in the inner details.

\subsection{Steep Nuclear Rise of Rotation and Implication on the Inner
Dynamics}

In our earlier paper (Sofue 1996, 1997) we have shown
that the steeply rising rotation curves are common to almost all
spiral galaxies of types Sb and Sc, including SBb and SBc,
for which high resolution CO data are available.
Such steeply-rising CO rotation in the central region has been also 
suggested for many southern galaxies (Bajaja et al 1995).
It  has been also found in many Virgo spirals
by the recent \ha\ spectroscopy (Rubin et al 1997).
This applies also in the present study, and our CCD \ha\ observations 
confirm these results.
This fact indicates that the mass distribution in the central
a few hundred parsecs is more steeply increasing toward the
nucleus than expected for an exponential-law surface-mass distribution.

By definition, a rotation curve is the trace of terminal
velocities in the position--velocity diagram along the major axis.
The assumption of axisymmetry and circular rotation has been extensively
adopted in deriving the mass of galaxies, including hypothetical dark halos.
The thus derived mass and potential have been used widely to discuss the 
dynamics of galactic disks, such as the resonance and density waves.
In many studies, however, the rotation curve for the central a few kpc
has been assumed to be rigid-body like.  

\noindent{\it Resonance}:
 If we adopt such steeply rising rotation curves in place of the widely assumed
rigid rotation, dynamical as well as magnetohydrodynamical analyses 
of the central  a few kpc region would be significantly affected. 
Since the epicyclic frequency, 
$ \kappa = \sqrt{1/2}~ \omega~ \sqrt{1-d~ {\rm ln} V(R) / d~{\rm ln} R}$,
gets minimum in a rigidly rotating disk, no resonance with the orbital rotation
occurs, where $\omega=V/R$ is the rotation frequency.
The resonance will occur at a radius, where the rotation curve turns
from rigid to flat, and the interestellar gas tends to be accumulated 
to make a ring of this turn-over radius.
On the other hand, if the rotation curve is not rigid but rather flat or
even declining, as often found in sharply peaking rotation curves,
this resonance ring occurs at a much smaller radius.
Hence, often quoted inner rings of radii a few hundre pc to a kpc
may not occur in such galaxies having sharply rising rotation curves.
In fact, many galaxies appear to have not necesarily a ring structure within
the central 1 kpc region: we can trace the outer spiral patterns 
continuously toward the very nuclear region.

\noindent{\it Dynamo}:
If we adopt the steeply-rising rotation curves,  some results on the 
galactic magnetic fields (e.g., Sofue et al 1986) based on the dynamo 
theory would be significantly changed for the central regions.
The dynamo action depends on the shearing motion of gas from a radius
to another.
If the rotation is rigid, the shearing motion is minimized, so that the
dynamo does not work, resulting in a weak or no magnetic fields in the inner
a few kpc region.
If the rotation is flat or declining, the shearing motion is more
significant and the dynamo action will be more effective, resulting in
a strong magnetic fields within the inner kpc region.

\subsection{Influence of Bars}

If a galaxy contains a bar, it is not straightforward to derive the 
mass distribution using the rotation curves.
Since dense molecular gas clouds and HII regions
are associated with shock-compressed gas lanes along the bar,
where the orbital velocity of gas is lowest, their staying probability
is highest along the shocked gas lane. On the other hand,
the probability is lowest when the gas clouds orbit in the inter-bar
regions, where the velocity is highest and nearly parallel to the bar.
Therefore, the observed gas clouds and HII regions are most
likely observed to exist in the shocked gas lane, which is rotating
nearly at the pattern speed of the bar.
This means that the apparent rotation velocities may be underestimated
compared to the circular velocity, if the galaxy has a bar.
Therefore, the presently discussed galaxies would have still steeper
nuclear rise of rotation, if they have bars.
Analysis of rotation curves taking into account
the effect of a bar is, however, beyond the scope of the present study:
the data are still too crude to discuss the bar, for which we
need to determine such a large number of parameters like
the mass of the bar, its three axes lengths, orientation of the
bar axes with respect to the line of sight, and the shocked gas motion.
Therefore, we stress that the present analysis of rotation curves
on the assumption of an axi-symmetry can give only a
first-order approximation of the mass distribution possibly still
under-estimated.

\subsection{Slowly-Rising Rotation}

{\it Dwarf and small-mass galaxies}:
We have seen that a large number of galaxies have steeply rising rotation
curve in the central a few kpc region.
These steep rises are seen both in H$_\alpha$ and in NII, as long 
as H$_\alpha$ is not absorbed significantly, as well as in CO.
However, there are some exceptions which show rather slowly rising rotation 
curves in CO, such as in the case of M33 (Sofue 1997).
In the present sample, NGC 1003, NGC 2403 and NGC 4605 have such rotation 
curves in \ha\ in the centeral regions.
As for NGC 2403, the CO rotation curve also shows a gentle slope in the 
central 5 kpc region.
Therefore, not every galaxy shows the nuclear steep rise.

From figures 3 and 4, one can see that galaxies with gentle rises have smaller 
disk rotation velocity than those with steep nuclear rises.
In fact, the maximum rotation velcity of NGC 2403 is about 130 km/s, 
and that of NGC 4605 is about 100 km/s, although it appears to be slightly 
rising at the observed outermost point.
On the other hand, galaxies with steep nuclear rise are likely to have 
the maximum rotation velocity of $\sim 200$ km/s or larger.
Also remarkable is that the galaxies with gentle rise do not have prominent 
bulges, and are classified as Sc or later.
Therefore, the slope of nuclear rotation curves are probably correlated 
with the galaxies' type and mass to some degree.
These facts may indicate that less massive galaxies like 
dwarfs, which have neither prominent bulge nor massive core,
are likely to show slowly-rising rotation in the nuclear regions,
whereas bright galaxies with disk rotation velocity of $\sim 150$ \kms\ or 
larger have steeply rising rotation curves, representing their massive bulge
and cores.

{\it Edge-on Galaxies}:
The central regions of edge-on galaxies are obscured by the foreground
dust in the disk: the observed rotation curves manifest the rotation
of a ring region at which the optical depth through the line of sight
becomes about unity.
In these cases, the central steep rise of rotation may not be observed, but
the derived curves show an apparently rigid rotation. 
Examples are seen for NGC 3495 and NGC 5907. NGC 4565 is also an edge-on galaxy,
while showing a central rise and peak, which may be significantly 
underestimated. 
In fact, CO-line observations, which does not suffer from interstellar 
absorption, shows a steep rise for these edge-on galaxies (Sofue 1997).

\acknowledgements{The observations were made at  the Okayama Astrophysical
Observatory of the National Astronomical Observatories.
We thank the staff of OAO for their help during the observations.
}


\newpage

\figcaption[fig1.eps]  
{[Left-upper panels]: $10' \times 10'$ image around the center
of each galaxy taken from the Digitized-Sky-Survey.
The slit position is indicated with the origin ($0'$) and
end ($4'$) of the slit, corresponding to those of the abscissa
in the other panels. 
[Left-lower panels]: Intensity profiles of the \ha\ and [NII]
lines in the form of 3D plot.
The abscissa is the relative displacement along the major axis
in arcseconds, and the ordinates the intensity in CCD counts.
[Right-upper and lower panels]:
[NII]- and \ha-line position-velocity diagrams in contour plots,
respectively.
The contours are drawn in logarithm of a base 2, with the lowest
contour level at 3 CCD counts, which is approximately 3 times
the rms noise.
The vertical dashed line indicates the center position of each
galaxy, which is the position of maximum intensity in the continuum
emission.
The horizontal line at the center indicates 2 kpc length
($\pm 1$ kpc) along the major axis, according to the distance
given in Table 1.
\label{fig1}}

\figcaption[fig2.eps]
{Velocity curves derived using the centroid of wavelengths
from the PV diagrams in the \ha\ and
[NII] line emssions.
The filled circles indicate data for \ha\ 6563,
and open circles for [N II] 6583 line.
The vertical dotted-line and the horizontal bar
have the same meaning as in the case of contour diagrams in figure 1.
\label{fig2}}

\figcaption[fig3.eps]
{Rotation curves derived from the PV diagrams using
the envelope-tracing method, plotted against the radius in kpc.
The horizontal bars indicate the angular length of 30$''$ on the sky.
\label{fig3}}

\figcaption[fig4.eps]
{The same as Fig. 3 but plotted in the same panel.
\label{fig4}}

\figcaption[fig5.eps]
{\ha\ and [NII] lines of NGC 3521 in a grey scale. This figure demonstrates
how the \ha\ Balmer absorption wing affects the PV diagram of the nuclear
region. In such a case, [NII] line is more safe to be used for
the rotation curve analysis.
\label{fig5}}

\newpage

\oddsidemargin -1cm
\evensidemargin -1cm
\begin{table*}
{\small{\small{\small{\small
\begin{center}
\caption{Parameters of observed galaxies}
\tablenum{1}
\begin{tabular}{cccccccccccccr}
\tableline
\tableline
Name & RA(1950) & Dec(1950)& $V_{\rm helio}$ & Size &
PA & $i$ & $B_T^{0,i}$ &  $D$ & $D_{\rm TF}$ & Type \\
& (h m s) &(\deg ~~$'$~~$''$) & (\kms) & ($'\times'$) &
(\deg) & (\deg) & (mag) & (Mpc) & (Mpc) &  \\
\tableline
NGC 1003& 02 36 06.12 & +40 39 28.0 &  627 &  5.5x 1.9 &  97 &
 67 & 11.54 & 10.7 & 11.6 & SA(s)cd        \\
NGC 1417& 03 39 28.20 &$-$04 51 50.0& 4068 &  2.7x 1.7 & 175 &
 50 & 12.26 & 54.1 & 52.1 & SAB(rs)b       \\
UGC03691& 07 05 10.50 & +15 15 33.0 & 2203 &  2.2x 1.1 &  65 &
 65 & 12.30 & 30.0 & 20.9 & SAcd           \\
NGC 2403& 07 32 05.50 & +65 42 40.0 &  131 & 21.9x12.3 & 127 &
 62 &  8.46 &  4.2 &  3.5 & SAB(s)cd       \\
NGC 2590& 08 22 28.50 &$-$00 25 42.0& 4998 &  2.2x 0.7 &  77 &
 71 & 12.99 & 64.5 & 70.0 & SA(s)bc        \\
NGC 2708& 08 53 36.90 &$-$03 10 05.0& 2008 &  2.6x 1.3 &  20 &
 68 & 12.55 & 24.6 & 55.0 & SAB(s)b; p?    \\
NGC 2841& 09 18 35.85 & +51 11 24.1 &  638 &  8.1x 3.5 & 147 &
 64 &  9.81 & 12.0 & 23.4 & SA(r)b         \\
NGC 2903& 09 29 20.30 & +21 43 23.9 &  556 & 12.6x 6.0 &  17 &
 66 &  9.13 &  6.3 &  8.7 & SAB(rs)bc      \\
NGC 3198& 10 16 51.94 & +45 48 06.0 &  663 &  8.5x 3.3 &  35 &
 71 & 10.55 & 10.8 & 11.5 & SB(rs)c        \\
NGC 3495& 10 58 40.90 & +03 53 43.0 & 1136 &  4.9x 1.2 &  20 &
 85 & 11.80 & 12.8 & 19.8 & Sd             \\
NGC 3521& 11 03 15.48 & +00 14 10.7 &  685 & 11.0x 5.1 & 163 &
 61 &  9.40 &  7.2 & 13.5 & SAB(rs)bc;L    \\
NGC 3672& 11 22 30.40 &$-$09 31 12.0& 1862 &  4.2x 1.9 &   8 &
 67 & 11.10 & 28.4 & 22.7 & SA(s)c         \\
NGC 4062& 12 01 30.50 & +32 10 26.0 &  769 &  4.1x 1.7 & 100 &
 68 & 11.66 &  9.7 & 18.8 & SA(s)c         \\
NGC 4321& 12 20 22.89 & +16 05 57.7 & 1571 &  7.4x 6.3 &  30 &
 37 & 10.00 & 16.8 & 14.7 & SAB(s)bc;HII   \\
NGC 4448& 12 25 46.10 & +28 53 50.0 &  669 &  3.9x 1.4 &  94 &
 71 & 11.50 &  9.7 & 24.4 & SB(r)ab        \\
NGC 4527& 12 31 35.17 & +02 55 43.5 & 1750 &  6.2x 2.1 &  66 &
 69 & 10.30 &  22    & 13.9 & Sb(c)          \\
NGC 4565& 12 33 52.00 & +26 15 47.0 & 1227 & 15.9x 1.9 & 136 &
 90 &  9.59 &  9.7 & 14.3 & SA(s)b;sp      \\
NGC 4569& 12 34 18.48 & +13 26 16.4 &$-$235&  9.5x 4.4 &  23 &
 69 &  9.86 & 16.8 & 10.3 & SAB(rs)ab;L;Sy \\
NGC 4605& 12 37 47.50 & +61 53 00.0 &  143 &  5.8x 2.2 & 125 &
 69 & 10.61 &  4.0 &  6.0 & SB(s)cp        \\
NGC 5033& 13 11 09.23 & +36 51 30.6 &  875 & 10.7x 5.0 & 170 &
 64 & 10.33 & 18.7 & 18.6 & SA(s)c         \\
NGC 5907& 15 14 35.90 & +56 30 45.4 &  667 & 12.8x 1.4 & 155 &
 90 & 10.31 & 14.9 & 18.0 & SA(s)c;sp      \\
NGC 7331& 22 34 46.66 & +34 09 20.9 &  821 & 10.5x 3.7 & 171 &
 68 &  9.67 & 14.3 & 16.9 & SA(s)b;L       \\
\tableline
\label{table1}
\end{tabular}
\end{center}
Note: L=LINER; Sy=Seyfert;  Distance $D$ is estimated by
$D=V_{\rm gc}/H_0$, where
$V_{\rm gc}$ is the galacto-centric distance and $H_0=75$ \kms/Mpc;
Distance $D_{\rm TF}$ was estimated by using a Tully-Fisher
relation from \cite{pie92}. All other parameters were taken
from NASA Extragalactic Data Base (NED).
}}}}
\end{table*}

\end{document}